\documentclass[a4paper,twoside,10pt]{article}
\usepackage{graphicx,publaob}

\def\papertype{Poster}
\begin{document}

\vfil

\title{NUMERICAL CODE FOR FITTING RADIAL EMISSION PROFILE OF A SHELL SUPERNOVA REMNANT}

\authors{B. ARBUTINA \lowercase{and} S. OPSENICA}


\address{Department of Astronomy, Faculty of Mathematics, University of Belgrade, Studentski trg 16, 11000 Belgrade, Serbia}
\Email{arbo}{matf.bg.ac}{rs} \Email{slobodan}{matf.bg.ac}{rs}

\markboth{RADIAL EMISSION PROFILE OF A SHELL SUPERNOVA REMNANT}{S. OPSENICA and B. ARBUTINA}

\abstract{Expressions for surface brightness distribution and for
flux density have been theoretically derived in the case of two
simple models of a shell supernova remnant. The models are: a
homogenous optically thin emitting shell with constant emissivity
and a synchrotron shell source with radial magnetic field.
Interactive Data Language (IDL) codes for fitting theoretically
derived emission profiles assuming these two models to mean
profiles of shell supernova remnants obtained from radio
observations have been written.}

\section{MODELS OF EMISSION OF SHELL SUPERNOVA REMNANTS}

\subsection{HOMOGENOUS OPTICALLY THIN EMITTING SHELL WITH CONSTANT EMISSIVITY}

In this paper, we investigated two models of emission of shell supernova remnants (SNRs). If we consider homogenous emitting shell with emissivity $\varepsilon _\nu$ = const, for specific intensity, if the medium is optically thin, we have
 \begin{equation}
 I_\nu  = \int  \varepsilon _\nu \mathrm{d}s = \Bigg\{ \begin{array}{ll}
 { \varepsilon _\nu (r_{2+}' - r_{1+}') + \varepsilon _\nu (r_{1-}' - r_{2-}'), } &  0< \sin \theta < \frac{R-\Delta}{d} \\
 { \varepsilon _\nu (r_{2+}' - r_{2-}') , }  &  \frac{R-\Delta}{d} \leq \sin \theta \leq \frac{R}{d},
 \end{array}
\end{equation}
where $\mathrm{d}s=\mathrm{d}r'$.
Cosine theorems (see Fig. 1)
 \begin{equation}
(R-\Delta)^2  = d^2 + r_1'^2 -2d r_1' \cos \theta ,
\end{equation}
 \begin{equation}
R^2  = d^2 + r_2'^2 -2d r_2' \cos \theta ,
\end{equation}
give us
 \begin{equation}
r_{1\pm}'  = d \cos \theta \pm \sqrt{(R-\Delta)^2 - d^2 \sin ^2 \theta} ,
\end{equation}
 \begin{equation}
r_{2\pm}'  = d \cos \theta \pm \sqrt{R^2 - d^2 \sin ^2 \theta} .
\end{equation}

Finally, we have
\begin{equation}
I_{\nu}=
\left\{
\begin{array}{ll}   C_{\nu}\left(\sqrt{\sin^{2}\theta_{2}-\sin^{2}\theta}-\sqrt{\sin^{2}\theta_{1}-\sin^{2}\theta}\right), & \mbox{ } 0 < \theta < \theta_{1} \\ C_{\nu}\sqrt{\sin^{2}\theta_{2}-\sin^{2}\theta}, & \mbox{ } \theta_{1} \leq \theta \leq \theta_{2},
\end{array}
\right.
\end{equation}
where $\theta_{1}=\arcsin\frac{R-\Delta}{d}$,
$\theta_{2}=\arcsin\frac{R}{d}$ and $C_{\nu}=2\varepsilon_{\nu}d$.
From the last equation one can see that $I_\nu ^0 = 2 \varepsilon
_\nu R \delta$ and $I_\nu ^{\mathrm{max}} = 2 \varepsilon _\nu R
\sqrt{\delta (2-\delta)} $, where $\delta = \Delta /R$. Brightness
distribution  i.e. specific intensity in units $2 \varepsilon _\nu
d$ given by equation (6), for $\delta = \Delta / R = 0.1$ and $R/d
= 0.01$, can be seen in Fig. 2.

\begin{figure}
\centerline{\includegraphics[bb= 0 0 576
288,width=\textwidth,keepaspectratio]{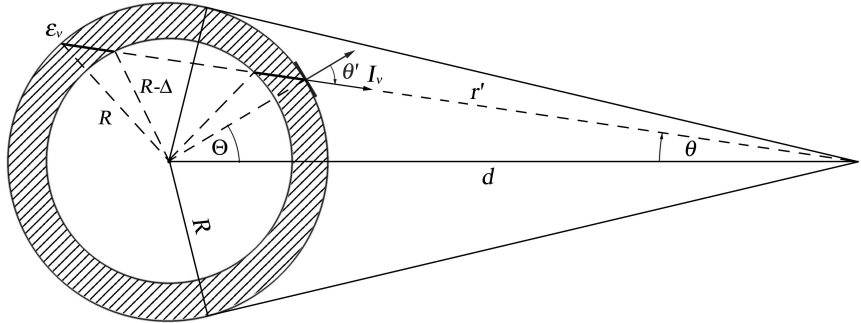}} \caption{
Radiation from an optically thin homogenous shell with thickness
$\Delta$ and radius $R$, at the distance $d$ from the observer.
$r_1'$ and $r_2'$ are the distances from intersection of the line
of sight for a given $\theta$ with the inner and outer radius of
the shell, respectively.}
\end{figure}

For the total flux density we have
\begin{eqnarray}
S_\nu &=& \int _0 ^{2\pi} \int _0 ^{\theta _s} I_\nu \cos \theta
\sin \theta \mathrm{d}\theta \mathrm{d}\varphi \nonumber \\
&=& 4\pi \varepsilon _\nu \int _0 ^{\theta _1} \Big( \sqrt{R^2 - d^2 \sin ^2 \theta} - \sqrt{(R-\Delta)^2 - d^2 \sin ^2 \theta } \Big) \cos \theta
\sin \theta \mathrm{d}\theta  \nonumber \\
&+& 4\pi \varepsilon _\nu \int _{\theta _1} ^{\theta _2}  \sqrt{R^2 - d^2 \sin ^2 \theta} \cos \theta
\sin \theta \mathrm{d}\theta .
\end{eqnarray}
After integration we obtain the expected result
\begin{equation}
S_\nu = \frac{4\pi}{3} \varepsilon _\nu d \Big[ \Big( \frac{R}{d} \Big)^3 -  \Big( \frac{R-\Delta}{d} \Big)^3 \Big] = \frac {\varepsilon _\nu V}{d^2} = \frac {\mathcal{E} _\nu V}{4\pi d^2} = \frac {L _\nu }{4\pi d^2},
\end{equation}
where the shell volume is $V = \frac{4\pi}{3}f R^3$, $f = 1-(1-\delta)^3$ is the volume filling factor, $\mathcal{E} _\nu = 4 \pi \varepsilon _\nu$ is total volume emissivity ($\varepsilon _\nu$ is emissivity per unit solid angle) and $L _\nu$ is luminosity.

\begin{figure}
\centerline{\includegraphics[bb= 0 0 308
228,width=0.9\textwidth,keepaspectratio]{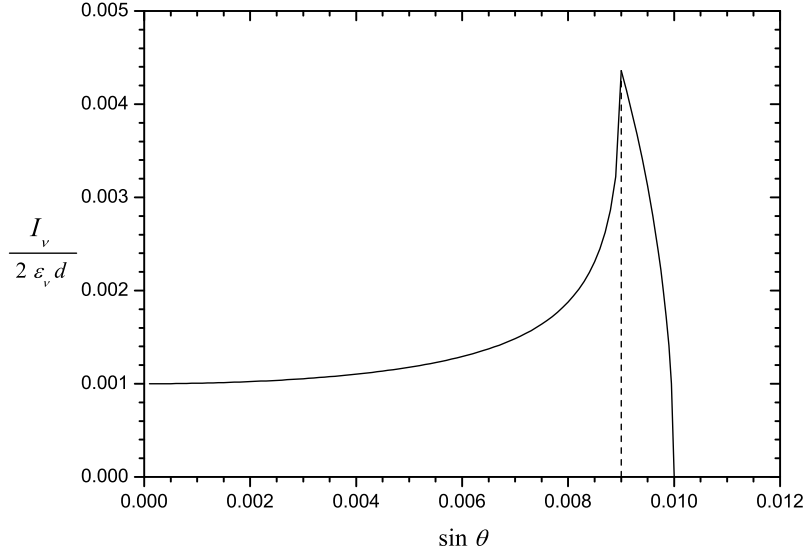}} \caption{
Brightness distribution for an optically thin homogenous
shell-like source with $\delta = \Delta / R = 0.1$ and $R/d =
0.01$.}
\end{figure}

\subsection{SYNCHROTRON SHELL SOURCE WITH RADIAL MAGNETIC FIELD}

If we have a synchrotron shell source with radial magnetic field,
emission coefficient is $\varepsilon _\nu \propto (B\sin \theta '
)^{\alpha +1} \nu ^{-\alpha}$ i.e.
\begin{equation}
\varepsilon _\nu = \tilde{\varepsilon}_\nu (\sin \theta '
)^{\alpha +1}.
\end{equation}
Sine theorem (see Fig. 1) gives us
\begin{equation}
\frac{r'}{d}=\frac{\sin \Theta}{\sin \theta'}, \ \ \ \Theta =
\theta ' - \theta ,
\end{equation}
i.e.
 \begin{equation}
r' = d (\cos \theta - \sin \theta \ \cot \theta ')
\end{equation}
and
 \begin{equation}
\mathrm{d}s=\mathrm{d}r' = d \sin \theta \frac{\mathrm{d}\theta '
}{\sin ^2 \theta '}
\end{equation}

 Intensity is then
 \begin{equation}
 I_\nu  = \int  \varepsilon _\nu \mathrm{d}s =    \tilde{\varepsilon} _\nu d \sin \theta \int (\sin \theta ')^{\alpha -1} \mathrm{d}\theta '
\end{equation}
i.e.
\begin{equation}
I_{\nu}=
\left\{
\begin{array}{ll}
2C_{\nu}\sin\theta\int^{\mu_{2-}}_{\mu_{1-}}\left(1-\mu^{2}\right)^{\frac{\alpha-2}{2}}d\mu, & \mbox{ } 0 < \theta < \theta_{1} \\ C_{\nu}\sin\theta\int^{\mu_{2-}}_{\mu_{2+}}\left(1-\mu^{2}\right)^{\frac{\alpha-2}{2}}d\mu, & \mbox{ } \theta_{1} \leq \theta \leq \theta_{2},
\end{array}
\right.
\end{equation}
where $\mu = \cos \theta '$, $\mu_{1,2\pm}=\mp\frac{\sqrt{\sin^{2}\theta_{1,2}-\sin^{2}\theta}}{\sin\theta_{1,2}}$ and $C_{\nu}=\tilde{\varepsilon_{\nu}}d$.

Rather than direct integration we will find flux density through
$
S_\nu =  \frac {L _\nu }{4\pi d^2}= \frac {\mathcal{E} _\nu
V}{4\pi d^2} ,
$
where
\begin{equation}
\mathcal{E} _\nu = \int_{4 \pi} \varepsilon _\nu \mathrm{d}\omega
' = \int_0^{2 \pi} \int _0^\pi \tilde{\varepsilon} _\nu (\sin
\theta ')^{\alpha +1} \sin \theta ' \mathrm{d}\theta '
\mathrm{d}\varphi = {2 \pi} \tilde{\varepsilon} _\nu \int _0^\pi
(\sin \theta ')^{\alpha +2}
 \mathrm{d}\theta '
\end{equation}
i.e. $\mathcal{E} _\nu = {2 \pi}  \sqrt{\pi}\frac{\Gamma
(\frac{\alpha +3}{2})}{\Gamma (\frac{\alpha +4}{2})}
\tilde{\varepsilon}_\nu $
 and the shell volume is as
before $V = \frac{4\pi}{3}f R^3$, $f = 1-(1-\delta)^3$.

\begin{figure}
\centerline{\includegraphics[bb= 0 0 308
228,width=0.9\textwidth,keepaspectratio]{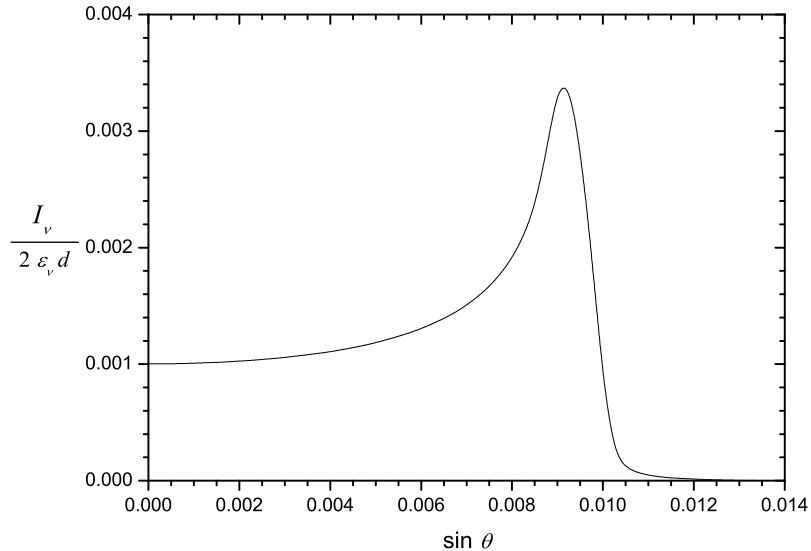}} \caption{
Radial profile with $\delta = \Delta / R = 0.1$ and $R/d = 0.01$
convolved with a Bessel function with HPBW=$5\times 10^{-4}$ rad.}
\end{figure}

\section{IDL CODES}

\subsection{DIRECT PROBLEM: SIMULATION OF OBSERVATIONS OF SHELL SUPERNOVA REMNANTS WITH RADIO TELESCOPE}

If we observe a shell SNR with a radio telescope, picture we get
is a convolution of real intensity of radiation of the SNR and
power pattern of the telescope, so we get "convolved" intensity
(Fig 3). When simulating this convolution numerically, one must
choose an expression for real intensity and an expression for
power pattern of a radio telescope. In our case, expressions for
intensities are (6) and (14), according to the models. For power
pattern $P_{n}\left(\theta\right)$, two possible cases have been
chosen: Gaussian approximation and approximation with Bessel
function of the first kind. Each pattern has defined half power
beam width (HPBW). Usually one takes HPBW=$1.02\frac{\lambda}{D}$
from Bessel function approximation ($D$ is diameter of radio
telescope and $\lambda$ is wavelength, see Rohlfs and Wilson 1996,
Uro\v sevi\'c and Milogradov-Turin 2007). Because of technical
limitations, we must consider that power pattern takes zero value
for angles greater than some critical angle $\theta_{c}$. In the
case of Gaussian approximation of power pattern, $\theta_{c}$ of
$5$ sigmas ($\sigma =\mathrm{HPBW}/(2\sqrt{\ln 2})$), while for
the approximation with Bessel function $\theta_{c}$ of $8$ HPBW
has been chosen.

Expression that is used for numerical simulation of convolution of
intensity of radio emission from a SNR and power pattern of a
radio telescope is:
\begin{equation}
I^{\mathrm{conv}}_{\nu}\left(\theta_{0}\right)=\frac{\int\int_{\mathrm{intersection}}
I_{\nu}\left(\theta\right)P_{n}(\theta^{'})\sin\theta d\theta
d\varphi}{2\pi\int^{\theta_{c}}_{0}P_{n}(\theta)\sin\theta
d\theta}.
\end{equation}
Angle $\theta^{'}$ is related to other angular parameters through
following relation of spherical trigonometry:
\begin{equation}
\cos\theta^{'}=\cos\theta\cos\theta_{0}+\sin\theta\sin\theta_{0}\cos\varphi.
\end{equation}
Region of double integration in numerator of the expression (16)
is the intersection between regions where two convolving functions
$I_{\nu}\left(\theta\right)$ and $P_{n}\left(\theta\right)$ are
defined. That double integration is performed by the IDL function
\texttt{INT\_2D}. Integration in denominator of the expression
(16) is performed by the IDL function \texttt{QROMB}. Integrations
in the expression (14) are performed by the IDL functions
\texttt{QROMB} and \texttt{INT\_TABULATED}, as well as by
"handwritten" function that calculates definite integrals using
rectangular method.

In the case of first model (with constant emissivity), user of the
program enters parameters $C_{\nu}$, $\theta_{1}$, $\theta_{2}$,
as well as parameter of antenna HPBW, and the program performs a
convolution. In the case of second model (with radial magnetic
field), user also enters an additional parameter of object -
spectral index $\alpha$.

\subsection{INDIRECT PROBLEM: FITTING MODEL TO OBSERVED PROFILE OF A SUPERNOVA REMNANT}

Indirect problem is the following: user enters observed radial
emission profile of a shell SNR in the form of a table, as well as
the parameter HPBW, and the program should find the best values
for parameters $C_{\nu}$, $\theta_{1}$ and $\theta_{2}$ in the
case of first model, or $C_{\nu}$, $\theta_{1}$, $\theta_{2}$ and
$\alpha$ in the case of second model, by fitting the chosen model
to the entered data. This is performed by the iterative IDL
procedure \texttt{CURVEFIT}. To perform this procedure, user has
to estimate initial values for the parameters. That can be done
using observed radial profile and the expression (6) or (14).
Initial value for spectral index $\alpha$ can be taken to be
$0.5$. This parameter is, however, better to be held fixed
(assuming that it is known from spectra). In addition to finding
the best values of the parameters, the program also calculates
their errors (i.e. standard deviations). Finally, the program
calculates the flux density of an SNR.

The program has been tested with artificially generated data.
Results of the application of the program to the real data will be
given elsewhere.

\bigskip

{\bf Acknowledgements.} During the work on this paper the authors
were financially supported by the Ministry of Education and
Science of the Republic of Serbia through the projects: 176004
'Stellar physics' and 176005 'Emission nebulae: structure and
evolution'.

\references

Rohlfs, K., Wilson, T. L.: 1996, \journal{Tools of Radio Astronomy}, Springer, Berlin.

Uro\v sevi\'c, D., Milogradov-Turin, J.: 2007, \journal{Teorijske osnove radio-astronomije}, Matemati\v cki fakultet, Beograd.

\end{document}